\documentclass[aps,prb,twocolumn,showpacs,amsmath,superscriptaddress,groupedaddress]{revtex4-1}
\usepackage{graphics}
\usepackage{epsfig}
\usepackage[colorlinks=true,citecolor=blue,linkcolor=red,linktocpage=true,pagebackref=false]{hyperref}
\usepackage{color,soul}

\begin{document}

\title{
Two-electron state from the Floquet scattering matrix perspective
}
\author{Michael Moskalets}
\email{michael.moskalets@gmail.com}
\affiliation{Department of Metal and Semiconductor Physics, NTU ``Kharkiv Polytechnic Institute", 61002 Kharkiv, Ukraine}

\date\today
\begin{abstract}
Two single-particle sources coupled in series to a chiral electronic waveguide can serve as a probabilistic source of two-particle excitations with tunable properties. 
The second-order correlation function, characterizing the state of emitted electrons in space-time, is expressed in terms of the Floquet scattering matrix of a source. 
It is shown that the Fourier transform of the correlation function, characterizing the emitted state in energy space, can be accessed with the help of an energy resolved shot-noise measurement. 
The two-electron state emitted adiabatically is discussed in detail.  
In particular, the two-electron wave function is represented via two different sets of single-particle wave functions accessible experimentally. 
\end{abstract}
\maketitle

\section{Introduction}
The realization of a high-speed on-demand single-electron source \cite{Blumenthal:2007ho,Feve:2007jx,Dubois:2013dv} has marked the birth of a new field focused on operations with electron wave-packets containing one to few particles propagating in a ballistic conductor. 
Inspired by quantum optics the several experiments demonstrating a single-particle nature of emitted electron wave-packets were reported.\cite{Mahe:2010cp,Bocquillon:2012if,Bocquillon:2013dp}  
The dynamical switching into different paths of individual electrons propagating ballistically was reported in Ref.~\onlinecite{Fletcher:2012te}. 
Provided the single-electron source is available, the engineering of few-electron  states becomes possible.     
Controlled emission of few electron wave-packets in mesoscopic conductor was already realized experimentally in Refs.~\onlinecite{Blumenthal:2007ho,Kaestner:2008gv,Fujiwara:2008gt,Leicht:2011ke,Kataoka:2011fu,Fricke:2013cc} using a dynamic quantum dot  and in Ref.~\onlinecite{Dubois:2013dv} using a voltage pulse with quantized flux as suggested in Refs.~\onlinecite{Levitov:1996ie,Ivanov:1997wz}. 

\begin{figure}[t]
\begin{center}
\includegraphics[width=80mm]{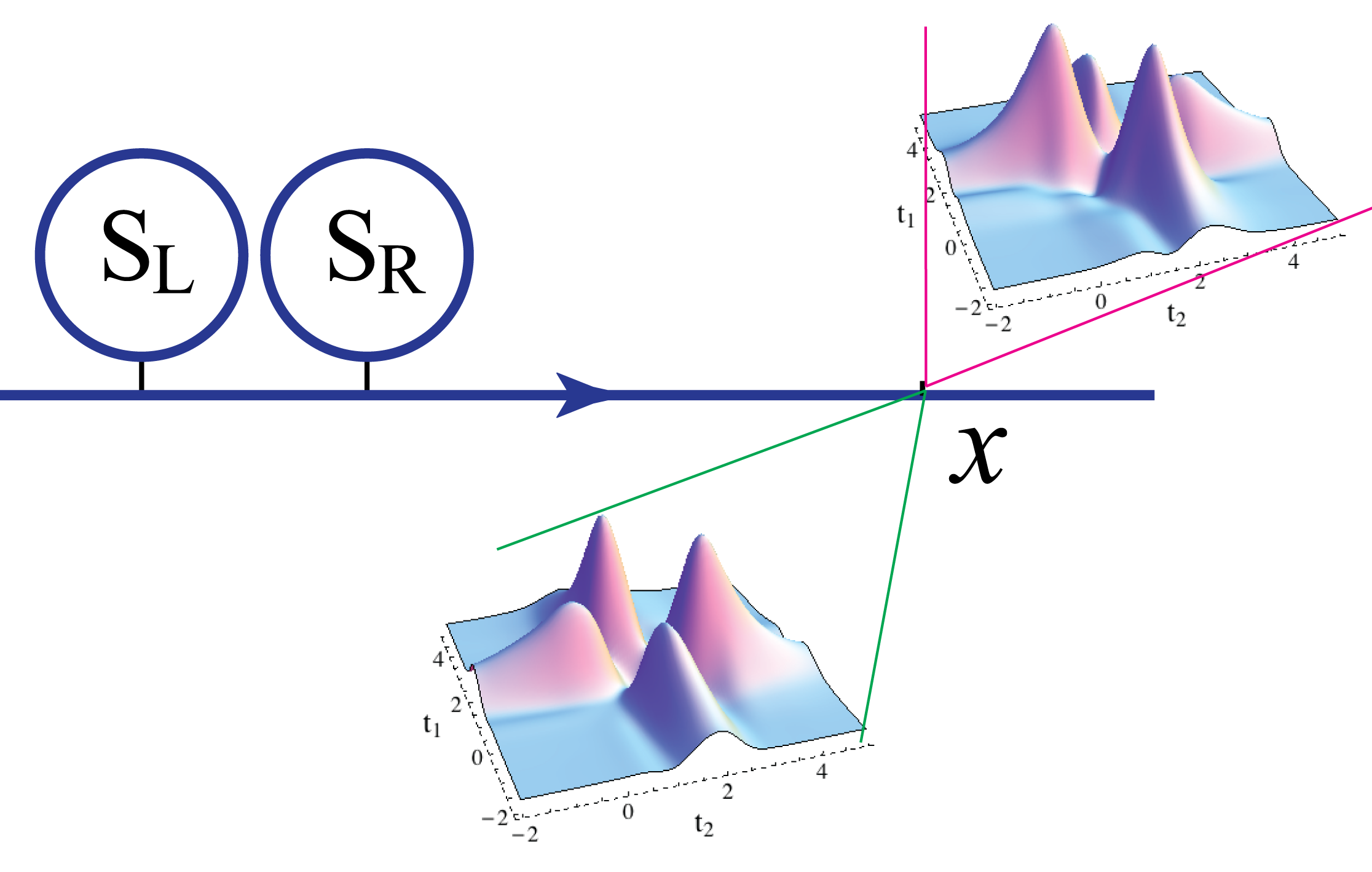}
\caption{
(Color online) Two single-electron emitters ${\rm S_{L}}$ and ${\rm S_{R}}$ attached to the same chiral electronic waveguide serve as a two-electron source if they emit electrons at close times $t_{L}^{-} \approx t_{R}^{-}$. {\bf Insets}: The real part of the envelope of a two-electron wave-function emitted adiabatically. It is shown at some position $x$ after the source and represented in different bases, $A_{L}^{(2)}(t_{1}x,t_{2}x)$ (upper inset) and $A_{R}^{(2)}(t_{1}x,t_{2}x)$ (lower inset), see Eq.~(\ref{15}). The parameters are following: emission times  $t_{L}^{-} = 2$, $t_{R}^{-} = 1$; coherence times of single-electron emitters $\Gamma_{R} = \Gamma_{L} = 1$. 
}
\label{source}
\end{center}
\end{figure}

The aim of this paper is to analyze a dynamical two-electron source composed of two periodically driven single-particle emitters attached to a chiral electronic waveguide, see Fig.~\ref{source}, as it is suggested in Ref.~\onlinecite{Splettstoesser:2008gc}. 
The advantage of such a two-particle emitter is a possibility to vary times when single electrons are emitted by individual sources and, therefore, continuously switch from the single-electron  emission to emission of pair of electrons. 
The close related source, addressed in Refs.~\onlinecite{Dubois:2013dv,Levitov:1996ie,Ivanov:1997wz,Keeling:2006hq,Dubois:2013fs,Lebedev:2005bt,Hassler:2009kw,Sherkunov:2009jm,Sherkunov:2012dg,Keeling:2008ft,Moskalets:2011jx,Vanevic:2008fn},  would utilize quantized Lorentzian voltage pulses  with variable center position.

In order to characterize the emitted two-particle state I extend the approach of Ref.~\onlinecite{Haack:2013ch} and introduce the second-order correlation function for emitted electrons. 
Within a scattering matrix formalism\cite{Moskalets:2011wz}, which describes well a single-particle source of Ref.~\onlinecite{Feve:2007jx}, see Ref.~\onlinecite{Parmentier:2012ed}, as well as the one used in Ref.~\onlinecite{Dubois:2013dv}, see Ref.~\onlinecite{Dubois:2013fs},  the source of electrons is described by the corresponding Floquet scattering matrix. 
I express the second-order correlation function in terms of the Floquet scattering matrix of the source and isolate the contribution due to emitted particles. 
This contribution can be represented as the Slater determinant composed of the first-order correlation functions, as it should be for non-interacting fermions\cite{Cahill:1999tn}, that justifies the decomposition into emitted electrons and electrons of the underlying Fermi sea we use. 
This procedure can be readily extended to describe an $n-$particle emitter and $n-$electron states.  
The close related approach to the first-order correlation function (single-electron coherence) is developed in Refs.~\onlinecite{Grenier:2011js,Grenier:2011dv} and applied to the analysis of $n-$electron Lorentzian pulses in Ref.~\onlinecite{Grenier:2013gg}. 
A Wigner function representation of the first-order electronic coherence is discussed in Ref.~\onlinecite{Ferraro:2013uv}.

The correlation function fully characterizes an emitted state. 
However presently it is a challenge  to access experimentally a correlation function on a single-electron level. 
On the other hand the first steps in this direction are already done. 
A time-resolved current profile on a single-electron level\cite{Feve:2007jx,Mahe:2008ut} and a single-electron wave-packet probability profile\cite{Fletcher:2012te,Dubois:2013dv} were reported. 
This inspires hope that the full quantum characterization of an emitted single to few-electron state, like it is done for single photons in optics\cite{Lundeen:2011hj,Polycarpou:2012kn}, is coming soon.

Another interesting object to look at is the energy distribution function, which is easier to access experimentally.
The non-equilibrium single-particle distribution function was already measured in Refs.~\onlinecite{Altimiras:2010ej,leSueur:2010fg,Altimiras:2010dk} via the energy resolved dc current and in Refs.~\onlinecite{Gabelli:2013ga,Dubois:2013dv} via the low-frequency shot noise spectroscopy. 
In this paper we discuss how to measure a two-particle distribution function via the energy resolved shot noise. 

Though the distribution function provides only partial information on the emitted two-particle state, nevertheless, it already demonstrates an essential feature of the state of two fermions propagating together, namely, an increase of the energy  compared to the case when they  propagate separately.\cite{Moskalets:2009dk,Moskalets:2013dl}$^{,}$\cite{Dubois:2013dv,Dubois:2013fs}  
We demonstrate this explicitly analyzing the evolution of the state emitted adiabatically. 
We calculate a two-particle wave function and show that with decreasing the time difference  between emission of two electrons it evolves from the product of single-electron wave functions to the Slater determinant composed of them. 
The single-electron wave functions in turn evolve from the bare ones emitted by the single-electron sources to the mutually orthogonal functions. 
The result of orthogonalization can be interpreted as if one bare single-electron wave function remains unchanged while the other one is adapted accordingly. 
Depending on which of two wave functions is kept unchanged, there are two bases for the representation of a two-electron wave function. 
Interestingly, performing measurements on the system with either one or two single-particle sources being switched on, one can access all four single-particle wave functions mentioned above. 

The paper is organized as follows:
In Sec.~\ref{tps} the second-order correlation function for emitted particles is calculated in terms of the Floquet scattering matrix of a periodically driven electron source. 
In Sec.~\ref{hmdf} we discuss how the Fourier transform of a correlation function, the energy distribution function, can be accessed via current cross-correlation measurement.    
In Sec.~\ref{ae} the two-electron state emitted adiabatically is analyzed in detail. 
A short conclusion is in Sec.~\ref{c}.

\section{General formalism}
\label{tps}

\subsection{Second-order correlation function for emitted particles}
\label{soc}

For a start let us define the second-order electronic correlation function, in the full analogy with how it is defined in optics,\cite{Glauber:2006cy} 

\begin{eqnarray}
{\cal G}^{(2)}\left(1,2,3,4  \right) =
\langle \hat\Psi^{\dag}(1) \hat\Psi^{\dag}(2) \hat\Psi(3) \hat\Psi(4) \rangle \,,  
\label{01}
\end{eqnarray}
\ \\
\noindent
where $\hat\Psi (j) \equiv \hat\Psi\left(x_{j}t_{j} \right)$ a single-particle electron field operator in second quantization evaluated at point $x_{j}$ and time  $t_{j}$,  $j=1,2,3,4$. 
The quantum-statistical average $\langle \dots \rangle$ is over the equilibrium state of electrons incoming to the source. 
To access information about emitted particles let us evaluate ${\cal G}^{(2)}$ after the source, where the field operator in second quantization for chiral electrons reads,\cite{Buttiker:1992vr}

\begin{eqnarray}
\hat\Psi\left(x_{j}t_{j} \right) = \int \frac{dE }{\sqrt{h v(E)} } e^{i\phi_{j}(E) } \hat b\left( E \right) \,.
\label{02}
\end{eqnarray}
\ \\
\noindent
Here $1/[h v(E)]$ is a one-dimensional density of states at energy $E$,  $\hat b(E)$ is an operator for electrons passed by (scattered off) the source, and the phase $\phi_{j}(E) = -E t_{j}/\hbar + k(E) x_{j}$.   

The electronic source driven periodically with frequency $\Omega$ is characterized  by the Floquet scattering matrix with elements $S_{F}\left( E_{n}, E \right)$, $E_{n} = E + n\hbar\Omega$, being amplitudes for an electron with energy $E$ to exchange $n$ energy quanta $\hbar\Omega$ with the scatterer.  
In such a case we can write,\cite{Moskalets:2002hu} 

\begin{eqnarray}
\hat b(E) = \sum_{n} S_{F}\left( E, E_{n} \right) \hat a\left( E_{n} \right)
\,, 
\label{ba}
\end{eqnarray}
\ \\
\noindent 
where $\hat a(E)$ is an operator for equilibrium electrons incoming to the scatterer. 
The average of the product  

\begin{eqnarray}
\left\langle \hat a^{\dag}(E) \hat a(E^{\prime}) \right\rangle = f_{0}(E)\delta\left( E - E^{\prime} \right)
\,,
\label{av}
\end{eqnarray}
\ \\
\noindent 
is defined by the Fermi distribution function $f_{0}(E) $ with temperature $T_{0}$ and the chemical potential $\mu$. 
Using quantities introduced above one can represent the correlation function ${\cal G}^{(2)}$ in terms of the Floquet scattering matrix of the source, 

\begin{eqnarray}
{\cal G}^{(2)}\left(1,2,3,4  \right) = \frac{1 }{2 }
\sum_{n,m,p,q} 
\iint 
\frac{dE dE^{\prime} f_{0}\left( E_{n} \right) f_{0}\left( E^{\prime}_{m} \right)  }{ h^{2} v( E ) v( E^{\prime} ) } 
\nonumber \\
\times 
S_{F}^{*}\left(E,E_{n}  \right)
S_{F}^{*}\left(E^{\prime},E^{\prime}_{m}  \right)
S_{F}\left(E_{p},E_{n}  \right)
S_{F}\left(E^{\prime}_{q},E^{\prime}_{m}  \right)
\nonumber \\
\times 
\det
\left ( 
\begin{array}{ll}
e^{i\phi_{1}(E)} 
&
e^{i\phi_{2}(E)} 
\\
e^{i\phi_{1}(E^{\prime})} 
&
e^{i\phi_{2}(E^{\prime})} 
\end{array}
 \right )^{*}
\det
\left ( 
\begin{array}{ll}
e^{i\phi_{4}(E_{p})} 
&
e^{i\phi_{3}(E_{p})} 
\\
e^{i\phi_{4}(E^{\prime}_{q})} 
&
e^{i\phi_{3}(E^{\prime}_{q})} 
\end{array}
 \right ).
\nonumber \\
\label{03}
\end{eqnarray}
\ \\
\noindent 
In above equation I use a wide band approximation and neglect the variation of the density of states on the scale of $\hbar\Omega$: $1/(h v\left( E_{n} \right) )\approx 1/(h v(E) )$. 
The structure of Eq.~(\ref{03}) tells us that the correlation function ${\cal G}^{(2)}$ is composed of elementary 2-particle propagators describing transfer of two electrons to points $4$ and $3$ from points $1$ and $2$. 
The action of the driven scatterer is to change energies of electrons, $E_{n}$ and $E^{\prime}_{m}$, as at destination points (to energies $E_{p}\,, E^{\prime}_{q}$) as at initial points (to energies $E\,, E^{\prime}$). 
The Fermi function $f_{0}\left( E_{n} \right)$ and $f_{0}\left(E^{\prime}_{m} \right)$ take care that the states with original energies, $E_{n}$ and $E^{\prime}_{m}$ are occupied.
To understand the content of ${\cal G}^{(2)}$ even more let us use the following identity,

\begin{eqnarray}
f_{0}\left( E_{n} \right) f_{0}\left( E^{\prime}_{m} \right) = 
f_{0}(E) f_{0}\left( E^{\prime} \right) 
\nonumber \\
+
f_{0}(E)\left [f_{0}\left(E^{\prime}_{m}\right)  - f_{0}(E^{\prime})  \right]
+
\left[ f_{0}\left(E_{n}\right)  - f_{0}(E)  \right] f_{0}\left( E^{\prime} \right)
\nonumber \\
+
\left [ f_{0}\left(E_{n}\right)  - f_{0}(E)  \right] \left [f_{0}\left(E^{\prime}_{m}\right)  - f_{0}(E^{\prime})  \right] 
\,.
\nonumber \\
\label{05} 
\end{eqnarray}
\ \\
\noindent
Four terms on the right hand side (RHS) of Eq.~(\ref{05}) results in four terms in ${\cal G}^{(2)}$, Eq.~(\ref{03}). 
The first term, $f_{0}(E) f_{0}\left( E^{\prime} \right)$, results in the second order correlation function for the Fermi sea incoming from the reservoir unperturbed by the driven scatterer. 
This is so, since the Floquet matrix elements drop out from the corresponding equation in force of the unitarity condition,\cite{Moskalets:2002hu} 

\begin{eqnarray}
\sum_{n}S_{F}^{*}\left(E,E_{n}  \right) S_{F}\left(E_{p},E_{n}  \right) = \delta_{p,0}
\,.
\label{uni}
\end{eqnarray}
\ \\
\noindent  
Next two terms results in contributions dependent only on two Floquet scattering elements and, therefore, can be interpreted as describing correlations between one unperturbed electron of the Fermi sea and one excited electron. 
And finally the last term on the RHS of Eq.~(\ref{05}) describes correlations between two excited electrons. 
This last contribution is of our interest here and we denote it as $G^{(2)}$. 
Therefore, the quantity $G^{(2)}$ is referred to as {\it the second-order correlation function for emitted particles}.   
As expected it can be represented as the determinant,   

\begin{eqnarray}
G^{(2)}\left(1,2,3,4  \right) = 
\det
\left ( 
\begin{array}{ll}
G^{(1)}(1,4)
&
G^{(1)}(1,3)
\\
G^{(1)}(2,4)
&
G^{(1)}(2,3)
\end{array}
 \right ) \,,
\label{g2}
\end{eqnarray}
\ \\
\noindent 
composed of the first-order correlation functions for emitted particles,\cite{Haack:2013ch}

\begin{eqnarray}
G^{(1)}(j,j^{\prime}) = 
\sum_{n,m=-\infty}^{\infty} 
\int 
\frac{dE  \left\{ f_{0}\left( E_{n} \right) - f_{0}(E)  \right\} }{ h v( E ) } 
\nonumber \\
\times 
S_{F}^{*}\left(E,E_{n}  \right)
S_{F}\left(E_{m},E_{n}  \right)
e^{-i\phi_{j}(E)} 
e^{i\phi_{j^{\prime}}(E_{m})} 
\,.
\label{06}
\end{eqnarray}
\ \\
\noindent
The fact that $G^{(2)}$ is expressed in terms of $G^{(1)}$ is a mere consequence of the well known Wick theorem. 
Note that $G^{(1)}$ is the first-order correlation function for the combined emitter.
It has no a simple relation to the states emitted by the single-particle sources working independently.

\subsection{Distribution functions for emitted particles}
\label{dffep}

To characterize the state of emitted particles in the energy space it is convenient to introduce the distribution function.

\subsubsection{Single-particle distribution function}

The single-particle distribution function for emitted particles is defined as follows: 

\begin{eqnarray}
f(E)\delta\left( E - E^{\prime} \right) = \left\langle \hat b^{\dag}(E) \hat b(E^{\prime}) \right\rangle\big|_{E^{\prime} = E} -  \left\langle \hat a^{\dag}(E) \hat a(E^{\prime}) \right\rangle
\,.
\nonumber \\
\label{06-1}
\end{eqnarray}
\ \\
\noindent 
It is a probability (density) that one can detect one particle in the state with energy $E$.  Accordingly to Eq.~(\ref{02}) such a state is a plane-wave state.  
In terms  of the Floquet scattering matrix of the source the single-particle distribution function reads,\cite{Moskalets:2002hu}

\begin{eqnarray}
f\left( E \right) = 
\sum_{n=-\infty}^{\infty} 
\left | S_{F}\left( E, E_{n} \right)  \right |^{2} 
\left\{ f_{0}\left( E_{n} \right) - f_{0}\left( E \right) \right\} 
 \,.
\label{singledistr}
\end{eqnarray}
\ \\
\noindent
The difference of Fermi functions entering above equation emphasizes that what calculated is related to excitations  not to the Fermi sea.   

Easy to see that $f\left( E \right)$ can also be calculated as the Fourier transform of the first-order correlation function for emitted particles $G^{(1)}(j,j^{\prime})$, Eq.~(\ref{06}), taken at $x_{j} = x_{j^{\prime}} \equiv x$, see also Ref.~\onlinecite{Grenier:2013gg}. 
Note $G^{(1)}$ is periodic in $t_{j^{\prime}}$ when the difference $t_{j} - t_{j^{\prime}}$ is kept constant.  
Therefore, performing a continuos Fourier transformation with respect to $\tau_{jj^{\prime}} =t_{j} - t_{j^{\prime}}$ and afterwards averaging over period ${\cal T} = 2\pi/\Omega$ the resulting function of $t_{j^{\prime}}$ we obtain a desired relation,  

\begin{eqnarray}
G^{(1)}\left( E \right) &=& 
\int\limits_{0}^{{\cal T}} \frac{d t_{j^{\prime}}  }{ {\cal T}}
\int\limits_{-\infty}^{\infty} d \tau_{jj^{\prime}} 
e^{-i\frac{E }{\hbar } \tau_{jj^{\prime}} } 
G^{(1)}(t_{j}x,t_{j^{\prime}}x) 
\nonumber \\
&=& 
\frac{f(E) }{v\left( E \right) }  
\,. 
\label{06-3} 
\end{eqnarray}

\subsubsection{Two-particle distribution function}

One can derive a similar equation relating a  two-particle distribution function for emitted particles $f(E,E^{\prime})$ and the Fourier transform of the second-order correlation function   $G^{(2)}(1,2,3,4)$ taken at $x_{1} = x_{4}$ and $x_{2} = x_{3}$. 
As before we perform a continuous Fourier transformation with respect to $\tau_{14} = t_{1} - t_{4}$ and $\tau_{23} = t_{2} - t_{3}$ and average over $t_{4}$ and $t_{3}$:  

\begin{widetext}

\begin{eqnarray}
 f\left( E, E^{\prime}; x_{1}, x_{2} \right) &=& 
v(E) v(E^{\prime})
 \int\limits_{0}^{{\cal T}} 
\frac{d t_{4}  }{ {\cal T}}
\int\limits_{0}^{{\cal T}} 
\frac{d t_{3}  }{ {\cal T}}
\int\limits_{-\infty}^{\infty} 
d \tau_{14}  e^{-i\frac{E }{\hbar } \tau_{14} }
\int\limits_{-\infty}^{\infty} 
d \tau_{23}  e^{-i\frac{E^{\prime} }{\hbar } \tau_{23} }   
G^{(2)}\left( t_{1}x_{1},t_{2}x_{2},t_{3}x_{2},t_{4}x_{1} \right)
\nonumber \\
&=& f(E) f(E^{\prime}) + \delta  f(E,E^{\prime};x_{1},x_{2})
\,. 
\label{06-4} 
\end{eqnarray}
where the irreducible part is 

\begin{eqnarray}
 \delta  f(E,E^{\prime};x_{1},x_{2})
= (-1)
\sum_{n=-\infty}^{\infty}
\left\{ f_{0}\left( E_{n} \right) - f_{0}\left( E \right) \right\} 
\sum_{m=-\infty}^{\infty}
\left\{ f_{0}\left( E^{\prime}_{m} \right) - f_{0}\left( E^{\prime} \right) \right\} 
\sum_{p=-\infty}^{\infty}
\sum_{q=-\infty}^{\infty}
e^{i x_{1}\left[ k(E^{\prime}_{q}) - k(E)  \right]}
\nonumber \\
\times 
e^{-i x_{2}\left[ k(E^{\prime}) - k(E_{p})  \right]}
\frac{ \left( \hbar\Omega/\pi \right)^{2} \sin^{2}\left( \pi \frac{E^{\prime} - E }{\hbar\Omega } \right) }{ \left( E^{\prime} - E + q \hbar\Omega \right) \left( E^{\prime} - E - p \hbar\Omega  \right) }
S_{F}^{*}\left(E,E_{n}  \right)
S_{F}\left(E_{p},E_{n}  \right)
S_{F}^{*}\left(E^{\prime},E^{\prime}_{m}  \right)
S_{F}\left(E^{\prime}_{q},E^{\prime}_{m}  \right)
.
\nonumber \\
\label{delf}
\end{eqnarray}
\ \\
\noindent
Here $E_{jn} = E_{j} + n\hbar\Omega$, $j = 1,2$. 
In general above equation is complex. 
However, if the energy difference is multiple of the energy quantum $\hbar\Omega$, i.e., $E^{\prime} - E = \ell\hbar\Omega$, $(\ell$ is an integer) then above equation becomes manifestly real and losses its dependence on spatial coordinates $x_{1}$ and $x_{2}$. 
Taking into account that now only $p = - q = \ell$ contribute to Eq.~(\ref{delf}), we arrive at the following,  

\begin{eqnarray}
f(E,E_{\ell}) &=& f(E) f(E_{\ell}) + \delta f\left( E,E_{\ell} \right)
\,,
\nonumber \\
\delta f(E,E_{\ell}) &=& (-1)
\bigg |
\sum_{n=-\infty}^{\infty}
\left\{ f_{0}\left( E_{n} \right) - f_{0}\left( E \right) \right\} 
S_{F}^{*}\left(E,E_{n}  \right)
S_{F}\left(E_{\ell},E_{n}  \right)
\bigg |^{2}
.
\label{delf1}
\end{eqnarray}
\ \\
\noindent
Now the quantity $f(E,E_{\ell})$ admits an interpretation as a joint detection probability to find one electron in the state with energy $E$ and the other electron in the state with energy $E_{\ell} = E + \ell \hbar \Omega$. 
For $\ell=0$ we find $f(E,E) = 0$. 
That is a consequence of the Pauli exclusion principle, according to which two electrons (fermions) cannot be in the same state, i.e., cannot have the same energy in our case.  
This feature becomes manifest if the distribution function is rewritten in terms of  determinants, 

\begin{eqnarray}
f(E,E_{\ell}) = 
\frac{1 }{2 } 
\sum_{n=-\infty}^{\infty}
\sum_{m=-\infty}^{\infty} 
\left\{ f_{0}\left( E_{n} \right) - f_{0}\left( E \right) \right\} 
\left\{ f_{0}\left( E_{m} \right) - f_{0}\left( E \right) \right\} 
\left | 
\det
\left ( 
\begin{array}{ll}
S_{F}\left(E,E_{n}  \right)
&
S_{F}\left(E,E_{m}  \right)
\\
S_{F}\left(E_{\ell},E_{n}  \right)
&
S_{F}\left(E_{\ell},E_{m}  \right)
\end{array}
 \right ) 
 \right |^{2}
 \,.
\label{slater}
\end{eqnarray}
\ \\
\noindent
Above equation is valid for arbitrary periodic driving. 
Similar equation but valid for adiabatic driving only was derived in Ref.~\onlinecite{Moskalets:2006gd}. 

Note namely the equation (\ref{delf1}) not (\ref{delf}) is relevant for measurable quantities  considered in this paper, see, e.g., Eq.~(\ref{noisedf}). 
\end{widetext}

\subsection{How to measure distribution functions for emitted particles}
\label{hmdf}

\begin{figure}[t]
\begin{center}
\includegraphics[width=80mm]{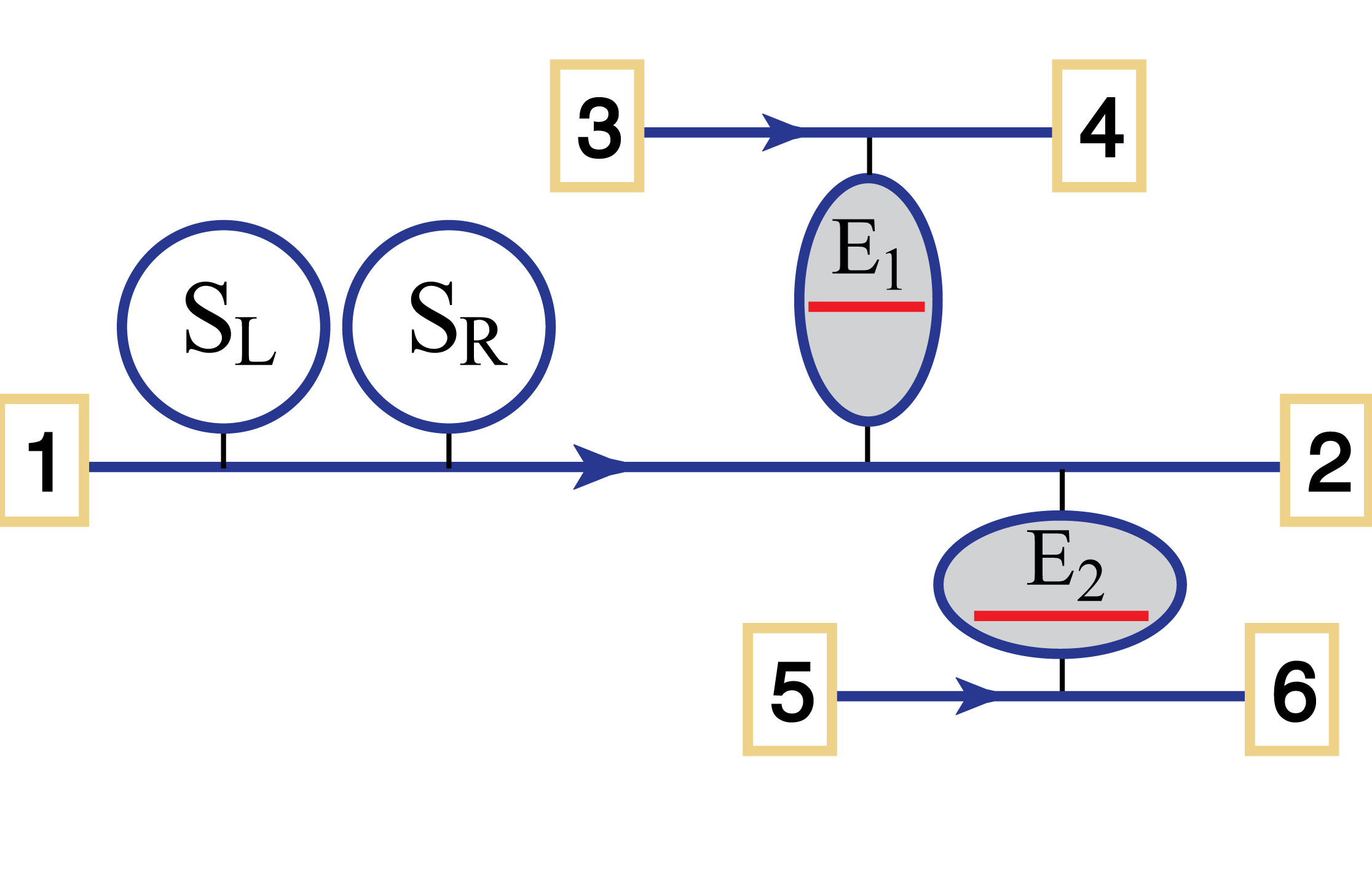}
\caption{
(Color online) To the measurement of a two-particle energy distribution function. $1$ through $6$ are metallic contacts. Blue lines with arrows are electronic chiral waveguides. Single-electron emitters ${\rm S_{L}}$ and ${\rm S_{R}}$ comprise a two-particle emitter. The quantum dots  with active levels ${\rm E_{1}}$ and ${\rm E_{2}}$ serve as energy filters. 
}
\label{set-up}
\end{center}
\end{figure}

The single-particle distribution function, as it was demonstrated experimentally,\cite{Altimiras:2010ej,leSueur:2010fg,Altimiras:2010dk} is related to the DC current through an energy filter, a quantum dot with a single conducting resonant level.
By analogy the two-particle distribution function can be accessed via the correlator of currents through two energy filters. 

The cartoon of a possible quantum coherent electronic circuit is shown in Fig.~\ref{set-up}. 
The two-particle source, composed of two single-electron sources ${\rm S_{L}}$ and ${\rm S_{R}}$, emits  electrons in pairs (possibly electrons and holes). 
All the metallic contacts $1$ through $6$ have the same potentials and the same temperatures. 
Electrons in contacts are in equilibrium and are characterized by the same Fermi distribution function $f_{0}(E)$.    
Blue solid straight lines are chiral electronic waveguides connecting metallic contacts. 
Such a waveguide can be, for instance, the edge state in the quantum Hall regime. 
The currents at the contacts $4$ and $6$ and their correlation function are of our interest.

The dynamical periodic source is characterized by the Floquet scattering amplitudes $S_{F}\left( E_{n},E \right)$, $E_{n} = E + n\hbar\Omega$, where $n$ is an integer, and $\Omega = 2\pi/{\cal T}$ with ${\cal T}$ the period of a drive.  
Two energy filters $\kappa = 1,2$, quantum dots, having one resonant level, $E_{\kappa}$, each  are attached to the central waveguide. 
The emitted electrons can escape to the contact $\beta = 4,6$ if they pass through the filter $\kappa = 1,2$, respectively.
   
To calculate a current, $I_{\beta}$, flowing to the contact $\beta = 4,6$, let us start from the  current operator in second quantization,\cite{Buttiker:1992vr}

\begin{eqnarray}
\hat I_{\beta}(t) &=&  \frac{e }{h } \iint dE dE^{\prime} e^{i t (E-E^{\prime})/h} 
\nonumber \\
&&\times
\{
\hat b_{\beta}^{\dag}(E)\hat b_{\beta}(E^{\prime}) 
- \hat a_{\beta}^{\dag}(E)\hat a_{\beta}(E^{\prime}) 
\}
\,.
\label{curop}
\end{eqnarray}
\ \\
\noindent
Note in the curly brackets above the first term is for particles entering the contact $\beta$ while the second one is for particles leaving the contact $\beta$. 
The path for the later particles is not shown in Fig.~\ref{set-up}.  

In the scattering matrix formalism the operators $\hat b$ for particles leaving the circuit to the contact $\beta$ are expressed in terms of operators $\hat a_{\gamma}$ for particles entering the circuit from the contact $\gamma$. 
For a dynamical circuit these operators are related via the Floquet scattering matrix of the circuit, $S_{F}^{cir}$:\cite{Moskalets:2002hu}

\begin{eqnarray}
\hat b_{\beta}(E) = \sum\limits_{\gamma} \sum\limits_{n=-\infty}^{\infty} S_{F,\beta\gamma}^{cir}\left( E,E_{n} \right) \hat a_{\gamma}(E_{n}) 
\,,
\label{ba1}
\end{eqnarray}
\ \\
\noindent
where $\gamma$ counts all the contacts where electrons can enter the circuit from. 
For the circuit shown in Fig.~\ref{set-up}, $\gamma = 1,3,5$. 
The relevant for us Floquet scattering matrix elements are, 

\begin{eqnarray}
S_{F,41}^{cir}\left( E,E_{n} \right) 
&=& 
e^{i\varphi_{41}(E)}
t_{1}(E) S_{F}(E,E_{n}) \,, 
\nonumber \\
S_{F,43}^{cir}\left( E,E_{n} \right) 
&=& 
e^{i\varphi_{43}(E)}
r_{1}(E) \delta_{n0}  ,
\nonumber \\
S_{F,61}^{cir}\left( E,E_{n} \right) 
&=&
e^{i\varphi_{61}(E)} 
t_{2}(E)r_{1}(E) S_{F}(E,E_{n}) \,, 
\nonumber \\
S_{F,65}^{cir}\left( E,E_{n} \right) 
&=& 
e^{i\varphi_{65}(E)} 
r_{2}(E) \delta_{n0}  \,,
\nonumber \\
S_{F,63}^{cir}\left( E,E_{n} \right) 
&=& 
e^{i\varphi_{63}(E)}
t_{2}(E)t_{1}(E) \delta_{n0}  ,
\nonumber \\
\label{floquet}
\end{eqnarray}
\ \\
\noindent
where 
$t_{\kappa}$/$r_{\kappa}$ is a transmission/reflection amplitude of the energy filter $\kappa = 1,2$, $\delta_{n0}$ is the Kronecker delta, $\varphi_{\beta\gamma}(E)$ is a phase of a free propagation through the circuit on the way from the contact $\alpha$ to the contact $\beta$. 
Since the circuit under consideration has no loops (it is single-connected) and it comprise only a single time-dependent source, the phases $\varphi_{\beta\gamma}$ are irrelevant.

\subsubsection{DC current}
\label{dcc}

The dc current flowing into the contact $\beta= 4,6 $ is calculated as a quantum statistical average over the (equilibrium) state of electrons incoming to the circuit, 

\begin{eqnarray}
I_{\beta} = \int\limits_{0}^{{\cal T}} \frac{dt }{{\cal T} } \left\langle \hat I_{\beta}(t) \right\rangle 
\,.
\label{dckappa}
\end{eqnarray}
\ \\
\noindent
To average we use Eqs.~(\ref{av}) and (\ref{ba1}) in Eq.~(\ref{curop}) and find,

\begin{eqnarray}
I_{\beta} = \frac{e }{h } \int dE F_{\beta}(E) \sum_{n=-\infty}^{\infty} \left | S_{F}\left( E,E_{n} \right) \right |^{2}
\nonumber \\
\times 
\left\{ f_{0}(E_{n}) - f_{0}(E) \right\}
\,,
\label{dc}
\end{eqnarray}
\ \\
\noindent 
where $F_{4}(E) = T_{1}(E)$ and $F_{6}(E) = R_{1}(E)T_{2}(E)$ with $T_{\kappa} = \left | t_{\kappa} \right |^{2}$, $\kappa = 1,2$ and $R_{1} = 1 - T_{1}$. 
Note while deriving above equation I used the relation, $t_{\kappa} r_{\kappa} = - t_{\kappa}^{*} r_{\kappa}^{*}$, which follows from the unitarity of the scattering matrices describing energy filters. 
Comparing Eq.~(\ref{dc}) and Eq.~(\ref{singledistr}) one can relate a dc current and the distribution function for emitted particles,\cite{Battista:2012db} 

\begin{eqnarray}
I_{\beta} = \frac{e }{h } \int dE F_{\beta}(E) f(E)
\,.
\label{dcdf}
\end{eqnarray}
\ \\
\noindent

\subsubsection{Zero frequency noise}
\label{zfn}

The zero-frequency cross-correlation function ${\cal P}_{46}$ for currents flowing to contacts $4$ an $6$ reads,\cite{Blanter:2000wi}

\begin{eqnarray}
{\cal P}_{46} = 
\int\limits_{0}^{{\cal T}} \frac{dt  }{{\cal T} } 
\int d t^{\prime} 
\frac{  
\left\langle 
\delta\hat I_{4}(t)  
\delta\hat I_{6}(t+t^{\prime})
+  
\delta\hat I_{6}(t+t^{\prime})
\delta\hat I_{4}(t)  
\right\rangle
 }{2 }
.
\nonumber \\
\label{noise}
\end{eqnarray}
\ \\
\noindent
Here $\delta\hat I_{\beta} = \hat I_{\beta} - \left\langle \hat I_{\beta} \right\rangle$ is an operator of current fluctuations. 
Let us use Eqs.~(\ref{curop}), (\ref{ba1}), and (\ref{floquet}) and calculate,

\begin{eqnarray}
{\cal P}_{46} = 
\frac{e^{2} }{2h }
\int dE
T_{1}(E)\Big[
- 
2 R_{1}(E) T_{2}(E)
\nonumber \\
\times
\sum_{n=-\infty}^{\infty}
\left\{ f_{0}\left( E_{n} \right) - f_{0}\left( E \right) \right\}^{2}
\left | S_{F}\left( E,E_{n} \right) \right |^{2}
\nonumber \\
+ 
\sum_{p=-\infty}^{\infty}
 R_{1}(E_{p}) T_{2}(E_{p})
\sum_{n=-\infty}^{\infty}
\sum_{m=-\infty}^{\infty}
\nonumber \\
\left\{ f_{0}\left( E_{n} \right) - f_{0}\left( E_{m} \right) \right\}^{2}
\nonumber \\
\times
S_{F}^{*}\left( E,E_{n} \right)
S_{F}\left( E,E_{m} \right)
S_{F}^{*}\left( E_{p},E_{m} \right)
S_{F}\left( E_{p},E_{n} \right)
\Big] \,.
\nonumber \\
\label{noise1}
\end{eqnarray}
\ \\
\noindent  
Since there is no a direct path between the contacts $4$ and $6$, the thermal noise does not appear in above equation.  

To simplify the equation above let us represent

\begin{eqnarray}
\left\{ f_{0}\left( E_{n} \right) - f_{0}\left( E_{m} \right) \right\}^{2} =
\left\{ f_{0}\left( E_{n} \right) - f_{0}\left( E \right) \right\}^{2}
\nonumber \\
+ \left\{ f_{0}\left( E_{m} \right) - f_{0}\left( E \right) \right\}^{2}
\nonumber \\
- 2 \left\{ f_{0}\left( E_{n} \right) - f_{0}\left( E \right) \right\}
\left\{ f_{0}\left( E_{m} \right) - f_{0}\left( E \right) \right\}\,.
\label{simp1}
\end{eqnarray}
\ \\
\noindent
One can use the unitarity of the Floquet scattering matrix, Eq.~(\ref{uni}), and show that the two first terms on the right hand side of Eq.~(\ref{simp1}) after substitution into Eq.~(\ref{noise1}) cancel  the first term on the right hand side of Eq.~(\ref{noise1}).
What remains is an equation of interest relating ${\cal P}_{46}$ and the irreducible part of a two-particle distribution function $\delta f$, Eq.~(\ref{delf1}):

\begin{eqnarray}
{\cal P}_{46} = 
\frac{e^{2} }{h }
\int dE
T_{1}(E)
\sum_{p=-\infty}^{\infty}
 R_{1}(E_{p}) T_{2}(E_{p}) \delta f(E,E_{p})
 .
\nonumber \\
\label{noisedf}
\end{eqnarray}
\ \\
\noindent  
Analogous equation but valid for adiabatic drive only was derived in Ref.~\onlinecite{Moskalets:2006gd}. 

Two equations (\ref{dcdf}) and (\ref{noisedf}) allow  to reconstruct a two-particle distribution function, $f(E,E^{\prime}) = f(E) f(E^{\prime}) + \delta f(E,E^{\prime})$, from all-electric measurements.    
To illustrate it let us consider ideal energy filters with $T_{\kappa}(E) = \gamma^{2}/\left( \left [ E - E_{\kappa} \right]^{2} + \gamma^{2}  \right)$.  
The width of the resonance $\gamma$ is assumed to be larger compared to the energy quantum $\hbar\Omega$ but smaller compared to the energy scale over which the Floquet scattering matrix changes.\cite{Battista:2012db} 
The later requirement is essential to keep $f$ in Eq.~(\ref{dcdf}) and $\delta f$ in Eq.~(\ref{noisedf}) at the resonant energies only. 
The former requirement allows us to simplify calculations and to replace $\sum_{n} \to \int d\omega_{n}/(\hbar\Omega)$ with $\omega_{n} = n\hbar\Omega$ and whenever necessary the Kronecker delta is replaced by the Dirac delta, $\delta_{nm} \to \hbar\Omega \delta\left(\omega_{n} - \omega_{m}  \right)$.

After simple calculations one can find, 
$I_{4} = e C_{1} f(E_{1})/{\cal T}$, 
$I_{6} = e C_{2} f(E_{2})/{\cal T}$, 
and 
${\cal P}_{46} = e^{2} C_{1}C_{2} \delta f(E_{1},E_{2})$, 
where $C_{1} = \pi \gamma/(\hbar\Omega)$ and $C_{2} = C_{1} (x^{2} + 2)/(x^{2} + 4)$ with $x = (E_{1} - E_{2})/\gamma$.  
Combining these equations together we finally arrive at the following relation,

\begin{eqnarray}
f\left( E_{1}, E_{2} \right) = \frac{{\cal T}^{2} }{e^{2} C_{1} C_{2} } \left\{ I_{4}(E_{1}) I_{6}(E_{2}) +\frac{{\cal P}_{46}(E_{1},E_{2}) }{{\cal T} } \right\} 
.
\nonumber \\
\label{df}
\end{eqnarray}
\ \\
\noindent
Above equation is derived in the case of a dynamical emission of electrons if all the contacts are grounded. 
In the stationary case but with biased contacts the analogous relation is widely used in mesoscopics, see, e.g., Refs.~\onlinecite{Chtchelkatchev:2002gp} and \onlinecite{Samuelsson:2006jw}.

\section{Adiabatic emission}
\label{ae}

As an example, showing how the general formalism developed above can be used to analyze  the emitted state, here I consider in detail the case when particles are emitted adiabatically into an electronic chiral waveguide. 
Such emission can be realized, for example, with the help of a slow driven mesoscopic capacitor\cite{Buttiker:1993wh,Gabelli:2006eg}, such as the one used in Ref.~\onlinecite{Feve:2007jx}, or using Lorentzian voltage pulses such as in Ref.~\onlinecite{Dubois:2013dv}. 
In the former case the source emits a stream of alternating electrons and holes, while in the later case an electron stream is emitted. 
The quantities related to adiabatic regime will be marked by the subscript ``$ad$''. 

I present results for a low temperature limit, 

\begin{eqnarray}
k_{B}T_{0} \ll \hbar\Omega_{0} \,.
\label{ltl}
\end{eqnarray}
\ \\
\noindent 
The generalization to finite temperatures is rather straightforward.

\subsection{Wave functions}
\label{esstr}

\subsubsection{First-order correlation function and two single-particle bases}

We calculate $G^{(1)}$, Eq.~(\ref{06}), for the source emitting particles (electrons or electrons and holes) adiabatically and denote it as $G^{(1)}_{ad}$. 
The adiabatic regime implies that the scattering amplitudes can be kept almost constant over the energy interval of order $\hbar\Omega$.\cite{Moskalets:2002hu} 
This allows us, first, to linearize the dispersion relation, for instance, $k\left( E_{n} \right) \approx k(E) + n\Omega/v(E)$. 
And, second, to calculate the Floquet scattering amplitude as the corresponding Fourier coefficient, 

\begin{eqnarray}
S_{F}\left( E_{n} , E \right) = S_{n}(E) \equiv \int\limits_{0}^{{\cal T}} \frac{dt }{{\cal T} } S(t,E) \, e^{i n \Omega t} 
\,,
\label{fourier}
\end{eqnarray}
\ \\
\noindent
of the frozen scattering amplitude, $S(t,E)$, which is the stationary scattering amplitude parametrically dependent on time. 
For low temperatures, $k_{B}T_{0} \ll \hbar\Omega$, let us make in Eq.~(\ref{06}) the following substitution, $f_{0}\left(E_{n}\right)  - f_{0}(E)  \approx \delta\left(E-\mu \right) n \hbar \Omega$,  and get,\cite{Haack:2011em}   

\begin{eqnarray}
G_{ad}^{(1)}(j,j^{\prime}) = 
\frac{i e^{-i \left [\phi_{j}(\mu)  -  \phi_{j^{\prime}}(\mu) \right] } }{2\pi v_{\mu} } 
\frac{ 1 - S^{*}\left( \tau_{j} \right) S\left( \tau_{j^{\prime}}\right)  }{\tau_{j} - \tau_{j^{\prime}} } 
\,. 
\label{07}
\end{eqnarray}
\ \\
\noindent
Here I introduced a reduced time $\tau_{j} = t_{j} - x_{j}/v_{\mu}$, denote $v_{\mu} \equiv v(\mu)$, and omit the energy argument, $S(\tau) \equiv S\left( \tau,\mu \right)$. 

Here we are interested in the regime when the source emits wave-packets comprising two particles. 
For definiteness we consider a source composed of two capacitors, ${\rm S_{L}}$ and ${\rm S_{R}}$, attached in series to the same chiral electronic waveguide, see Fig.~\ref{source}, and emitting particles at close times, $t^{-}_{L}$ and $t^{-}_{R}$, respectively. 
To be precise, let us concentrate on a two-electron emission. 
A two-hole emission can be analyzed in the same way. 
An electron-hole pair emission in the adiabatic regime is trivial\cite{Keeling:2006hq} and we do not address  it here.  
In particular, in the case of two identical capacitors there is the reabsorption effect:\cite{Splettstoesser:2008gc,Moskalets:2013dl} an electron (a hole) emitted by the first capacitor is reabsorbed by the second capacitor attempting to emit a hole (an electron) at the same time. 
As a consequence nothing is emitted. 
In contrast, in the non-adiabatic regime an electron-hole pair is emitted.\cite{Moskalets:2013dl}

The close related case of $n-$particle Lorentzian wave-packets is discussed in detail in Ref.~\onlinecite{Dubois:2013fs,Grenier:2013gg}.

In the adiabatic regime the scattering amplitude of the entire source is the product of scattering amplitudes of its constituents, capacitors, $S(\tau) = S_L(\tau) S_R(\tau)$.  
Close to the time of emission of an electron, $t_{\alpha}^{-}$, the scattering amplitude of the  capacitor $\alpha$ can be represented in the Breit-Wigner form,\cite{Olkhovskaya:2008en,Keeling:2008ft}

\begin{eqnarray}
S_{\alpha}(\tau) = \frac{\tau - \tau^{-}_{\alpha} + i \Gamma_{\alpha}}{\tau - \tau^{-}_{\alpha} - i \Gamma_{\alpha} } \,.
\label{08}
\end{eqnarray}
\ \\
\noindent 
Here $\tau^{-}_\alpha = t^{-}_{\alpha} - x_{\alpha}/v_{\mu}$ is the reduced emission time with $x_{\alpha}$ the coordinate of the source $\alpha$; 
$\Gamma_{\alpha} \ll {\cal T}$ is the half-width of the density profile\cite{Olkhovskaya:2008en} and, correspondingly, the coherence time\cite{Haack:2011em,Haack:2013ch} of the single-electron state emitted by the capacitor $\alpha = L, R$. 
Above equation is given for a single period, $0 < \tau < {\cal T}$. 
To other times it should be extended periodically, $S_\alpha(\tau) = S_\alpha(\tau + {\cal T})$.

\paragraph{Single-electron emission.} 

If each capacitor would work independently then it would emit an electron on the top of the Fermi sea in the state with the following wave function

\begin{eqnarray}
\Phi_{\alpha}(j) \equiv \Phi_{\alpha}(x_{j}t_{j}) &=& A_{\alpha}\left( \tau_{j} \right) e^{-i  \phi_{j}(\mu) } \,,
\nonumber \\ 
A_{\alpha}\left( \tau_{j} \right) &=& 
\sqrt{\frac{\Gamma_{\alpha} }{\pi v_{\mu} }}
\frac{1 }{ \tau_{j} -\tau^{-}_\alpha - i\Gamma_{\alpha}  } 
\,.
\label{09}
\end{eqnarray}
\ \\
\noindent 
The wave function given above can be inferred from the first-order correlation function, Eq.(\ref{07}), with $S=S_{\alpha}$, which is now factorized,\cite{Grenier:2013gg} 

\begin{eqnarray}
G_{ad,\alpha}^{(1)}(j,j^{\prime}) = \Phi^{*}_{\alpha}(j) \Phi_{\alpha}(j^{\prime} ) 
\,.
\label{single}
\end{eqnarray}
\ \\
\noindent
See also Ref.~\onlinecite{Keeling:2006hq} for alternative derivation. 

Straightforward calculations show that the corresponding second-order correlation function, see Eq.~(\ref{g2}), is zero, $G^{(2)}_{ad} = 0$, witnessing  a single-particle emission. 
Note the times $\tau_{j}$ and $\tau_{j^{\prime}}$ are within the same period, such that the particles emitted at different periods do not contribute to $G^{(2)}_{ad}$  

\paragraph{Two-electron emission.}

To calculate $G_{ad}^{(1)}$ for a two-particle source with $S(\tau)=S_{L}(\tau)S_{R}(\tau)$, I use in Eq.(\ref{07}) the following identity $1-ab = 0.5[1-a][1+b] + 0.5[1-b][1+a]$ with $a=S_{L}^{*}(\tau_j)S_L(\tau_j^\prime)$, $b=S_{R}^{*}(\tau_j)S_R(\tau_j^\prime)$ and find

\begin{eqnarray}
G_{ad}^{(1)}(j,j^{\prime}) = \frac{1}{2} \sum\limits_{\alpha = L,R} 
\left\{
 \Phi^{*}_{\alpha}(j) \Phi_{\alpha}(j^{\prime}) + \Phi^{*}_{\alpha\bar\alpha}(j) \Phi_{\alpha\bar\alpha}(j^{\prime}) 
\right\}
,
\nonumber \\
\label{10} 
\end{eqnarray}
\ \\
\noindent
where $\bar\alpha = L (R)$ for $\alpha = R (L)$ and 

\begin{eqnarray}
\Phi_{\alpha\bar\alpha}(j) \equiv \Phi_{\alpha\bar\alpha}(x_{j}t_{j}) &=& A_{\alpha\bar\alpha}\left( \tau_{j} \right) e^{-i  \phi_{j}(\mu) } \,,
\nonumber \\ 
A_{\alpha\bar\alpha}\left( \tau_{j} \right) &=& S_{\alpha}(\tau_{j}) A_{\bar\alpha}(\tau_{j})
\,.
\label{11}
\end{eqnarray}
\ \\
\noindent 
Note the functions $\Phi_{\alpha}$ and $\Phi_{\alpha\bar\alpha}$ (correspondingly, the envelop functions $A_{\alpha}$ and $A_{\alpha\bar\alpha}$) are mutually orthogonal, 

\begin{eqnarray}
\int dx \Phi_{\alpha}(xt)\Phi_{\alpha\bar\alpha}^{*}(xt) = 0\,,
\label{12}
\end{eqnarray}
\ \\
\noindent 
and normalized. 
Therefore, they can serve as a basis for the representation of a two-particle state of emitted electrons. 
Note that this basis is time-dependent. 
The unitary rotation from one basis to the other is also time-dependent.  
As we will see later on, this results in a basis-dependent energy distribution for each of electrons. 
While the energy distribution for two electrons is basis-independent.  

Let us choose the basis corresponding to some $\alpha$. 
Then representing $\Phi_{\bar\alpha}$ and $\Phi_{\bar\alpha\alpha}$ in terms of the basis functions $\Phi_{\alpha}$ and $\Phi_{\alpha\bar\alpha}$ one can rewrite Eq.~(\ref{10}) as the sum of two terms, 

\begin{eqnarray}
G_{ad}^{(1)}(j,j^{\prime}) =  \Phi^{*}_{\alpha}(j) \Phi_{\alpha}(j^{\prime}) + \Phi^{*}_{\alpha\bar\alpha}(j) \Phi_{\alpha\bar\alpha}(j^{\prime}) 
\,.
\label{double} 
\end{eqnarray}
\ \\
\noindent
Note here $\alpha = L$ or $\alpha = R$. 
There is no a summation over $\alpha$ on the right hand side in above equation.

The two terms on the right hand side of Eq.~(\ref{double}) are single-particle propagators for one  electron emitted in the state with the wave function $\Phi_{\alpha}$ and the other one with the wave function  $\Phi_{\alpha\bar\alpha}$, respectively.  
The close related representation but for a pulse comprising $n$ identical (with the same $\Gamma$ and emitted at the same time) particles is given in Ref.~\onlinecite{Grenier:2013gg}.

If in a waveguide the electrons propagate to the right, see Fig.~\ref{source}, then it is natural to choose the basis corresponding to $\alpha = R$, i.e., the basis functions are $\Phi_{R}$ and $\Phi_{RL}$.
Then the equation above admits an intuitive and transparent interpretation.   
An electron in the state with a  wave function $\Phi_{R}(xt)$ is emitted by the rightmost single-electron source  and  propagates away. 
Another electron is emitted by the leftmost single-electron source and it passes by the second source. 
If the later source would not work the wave function of a second electron  would be $\Phi_{L}(xt)$ (times an irrelevant constant phase factor  $S_{R}(\tau) = {\rm const}$, $|S_{R}|^{2} = 1$). 
However the working second source adds an extra non-trivial (i.e., time-dependent) factor $S_{R}(\tau)$ to the wave function of an electron passing it.\cite{Juergens:2011gu} 
Therefore, the corresponding  wave function becomes $\Phi_{RL}(xt) = S_{R}(\tau) \Phi_{L}(xt)$, $\tau = t-x/v_{\mu}$. 
The extra factor $S_{R}(\tau)$  in $\Phi_{RL}$ is responsible for the orthogonalization of single-particle states that is necessary for two electrons (fermions) to propagate in close vicinity to each other. 
An intuitive interpretation presented here is possible due to a properly chosen single-particle wave function basis. 
If we would use another basis the interpretation would be less transparent, while the description would be still correct. 

\begin{figure}[t]
\begin{center}
\includegraphics[width=80mm]{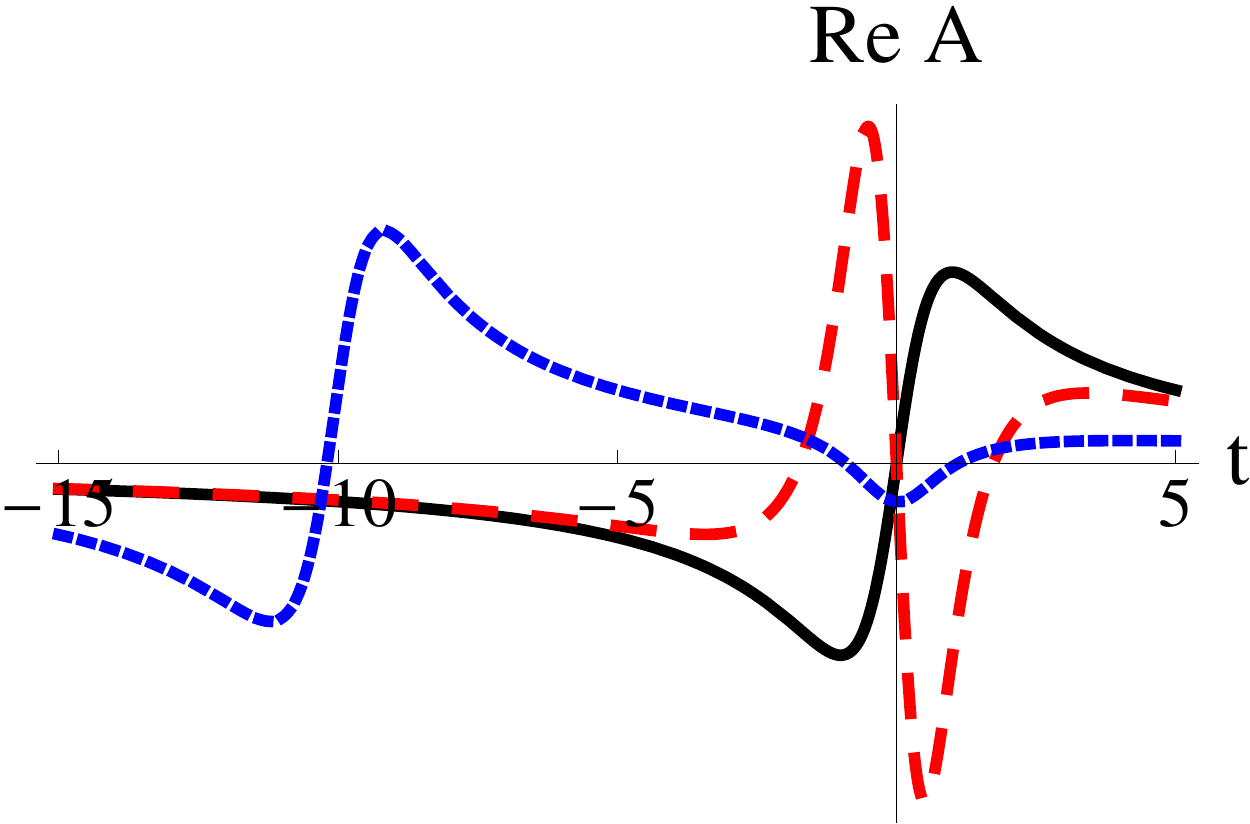}
\quad \\
\includegraphics[width=80mm]{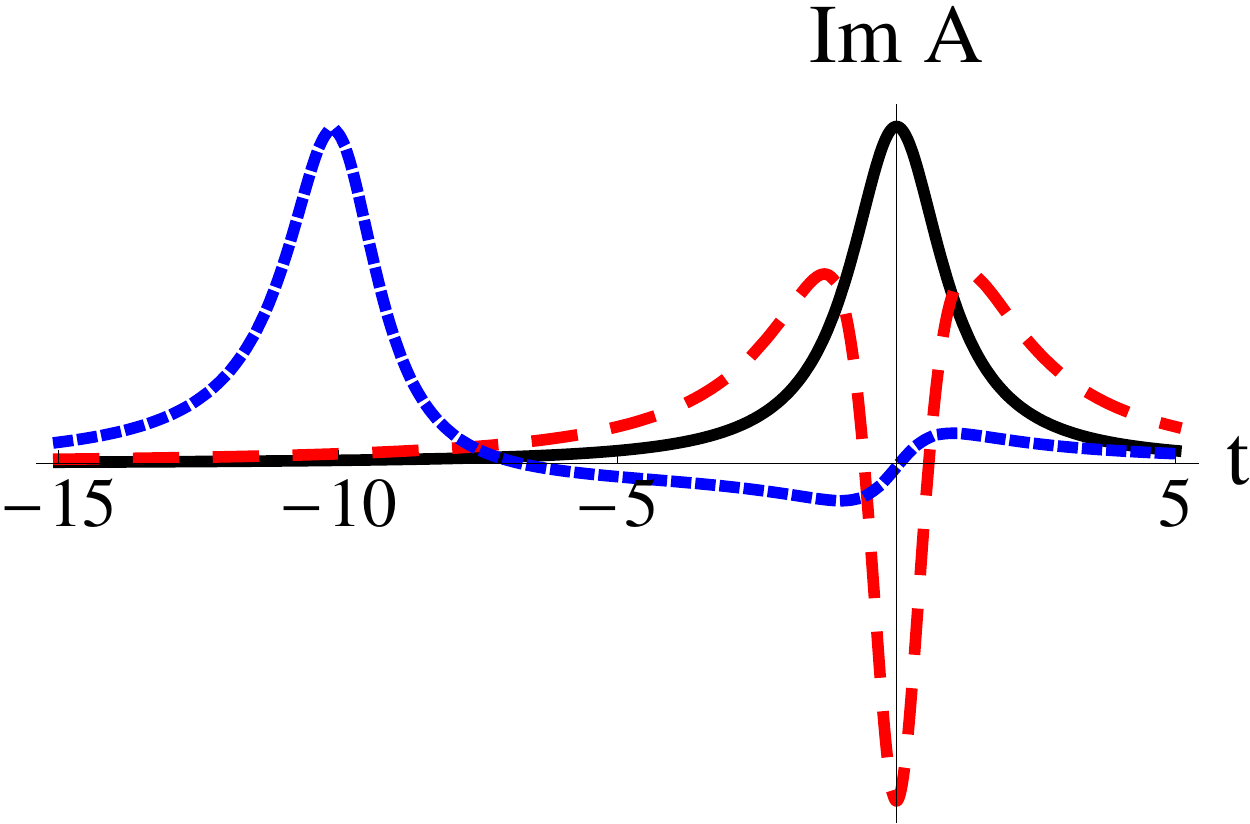}
\caption{
(Color online) Single-particle wave functions for electrons comprising a pair. 
The real part (upper panel) and imaginary part (lower panel) of the envelope functions $A_{R}$, Eq.~(\ref{09}), (black solid line) and $A_{RL}$, Eq.~(\ref{11}),  are shown as a function of time. The earlier times correspond to events happening first. At simultaneous emission, $t_{L}^{-} =  t_{R}^{-}$, $A_{RL}$ (red dashed line) is quite different from $A_{R}$. At successive emission,  $t_{R}^{-} - t_{L}^{-} = 10\Gamma_{R}$, $A_{RL}$ (blue dotted line) and $A_{L}$ are essentially the same. The parameters are following: emission time  $t_{R}^{-} = 0$; coherence times of single-electron emitters $\Gamma_{R} = \Gamma_{L} = 1$. 
}
\label{wf}
\end{center}
\end{figure}

The effect of one source on the electron emitted by the other source depends essentially on the difference of times, $\Delta\tau = \tau^{-}_{R} - \tau^{-}_{L}$, when electrons are emitted.  
As an illustration in Fig.~\ref{wf} the envelope functions $A_{R}$ (black solid line) and $A_{RL}$ are contrasted in the case of equal coherence times, $\Gamma_{L} = \Gamma_{R} \equiv \Gamma$. 
If the two electrons are emitted at the same time, $\Delta\tau = 0$, the envelope function $A_{RL}$ (red dashed line) differs substantially from $A_{R}$. 
While if the two electrons are emitted with a long time delay, $\Delta\tau = 10\Gamma$, the envelop function $A_{RL}$ (blue dotted line) resembles essentially $A_{R}$.

\subsubsection{Two-particle wave function}

Substituting Eq.~(\ref{double}) into Eq.~(\ref{g2}) one can factorize the second-order correlation function, 

\begin{eqnarray}
G^{(2)}_{ad}\left(1,2,3,4  \right) = \left( \Phi^{(2)}_{\alpha}(1,2)  \right)^{*} \Phi^{(2)}_{\alpha}(4,3) \,,
\label{14}
\end{eqnarray}
\ \\
\noindent
and correspondingly find a two-particle wave-function, 

\begin{eqnarray}
\Phi^{(2)}_{\alpha}(j,j^{\prime}) &=& 
A^{(2)}_{\alpha}(\tau_{j},\tau_{j^{\prime}}) e^{-i  \left [ \phi_{j}(\mu)  +  \phi_{j^{\prime}}(\mu)\right]} \,,
\nonumber \\ 
A^{(2)}_{\alpha}(\tau_{j},\tau_{j^{\prime}}) &=& 
\det \left ( 
\begin{array}{ll}
A_{\alpha}(\tau_{j})
&
A_{\alpha\bar\alpha}(\tau_{j})
\\
A_{\alpha}(\tau_{j^{\prime}})
&
A_{\alpha\bar\alpha}(\tau_{j^{\prime}})
\end{array}
 \right ) \,.
 \label{15}
\end{eqnarray}
\ \\
\noindent 
Remind that the reduced time is $\tau = t - x/v_{\mu}$. 
This wave function is the Slater determinant composed of single-particle wave-functions $\Phi_{\alpha}$ and $\Phi_{\alpha\bar\alpha}$ constituting the basis.\cite{Grenier:2013gg} 
Therefore, the Pauli exclusion principle for fermions is satisfied manifestly, $\Phi^{(2)}_{\alpha}(j,j) = 0$. 

Note depending on the base wave functions chosen, $\alpha = L$ or $\alpha = R$, the time-profile of a two-particle wave function $\Phi^{(2)}_{\alpha}$ will be different, see insets to Fig.~\ref{source}, while the propagator $G_{ad}^{(2)}$, Eq.~(\ref{14}), remains the same. 
Note also that the two-particle density profiles is basis-independent, $\left | \Phi^{(2)}_{L}(j,j^{\prime}) \right |^{2} = \left | \Phi^{(2)}_{R}(j,j^{\prime}) \right |^{2}$. 

With increasing the difference of emission times, $\Delta\tau = \tau^{-}_{R} - \tau^{-}_{L} \gg \Gamma_{\Sigma}$, the two-particle wave function $\Phi^{(2)}_{\alpha}(j,j^{\prime})$ is noticeable only for $\tau_{j} \approx \tau_{L}^{-}$, $\tau_{j^{\prime}} \approx \tau_{R}^{-}$, or for $\tau_{j} \approx \tau_{R}^{-}$, $\tau_{j^{\prime}} \approx \tau_{L}^{-}$. 
In any of these cases the determinant in Eq.~(\ref{15}) has only two non-zero entries, either along the main diagonal or the other two. 
Apparently the two-particle state is now the product of two single-particle states.

\subsubsection{Two bases: How to measure}
\label{tbhtm}

One of the possibility to access bases wave functions is to measure the first-order correlation function and utilize its additivity property, see Eq.~(\ref{double}). 
Two protocols, a single-electron quantum tomography\cite{Grenier:2011js,Grenier:2011dv}  and a time-resolved single-electron interferometry\cite{Haack:2011em,Haack:2013ch},  are already proposed for such a measurement. 
 
First, let us switch on only one single-particle source. 
Then, accordingly to Eq.~(\ref{single}), if only the capacitor $S_{L}$ works, we measure  a single-particle propagator $G_{ad,L}^{(1)}$ and derive the wave function $\Phi_{L}$. 
In the same way, if only the capacitor $S_{R}$ works, we measure $G_{ad,R}^{(1)}$ and obtain the wave-function $\Phi_{R}$.  
As the next step let us perform a measurement with two single-particle sources switched on.
The corresponding measurement provides us with the first-order correlation function, $G_{ad}^{(1)}$, Eq.~(\ref{double}). 
This correlation function can be represented either as $G_{ad}^{(1)} = G_{ad,L}^{(1)} + G_{ad,LR}^{(1)}$ or as $G_{ad}^{(1)} = G_{ad,R}^{(1)} + G_{ad,RL}^{(1)}$. 
Therefore, combining the latter measurement with one of the former measurements one can  derive $G_{ad,RL}^{(1)}(j,j^{\prime}) = \Phi_{RL}^{*}(j)\Phi_{RL}(j^{\prime})$ and $G_{ad,LR}^{(1)}(j,j^{\prime}) = \Phi_{LR}^{*}(j)\Phi_{LR}(j^{\prime})$, correspondingly. 
Combining together $\Phi_{R}$ and $\Phi_{RL}$ or $\Phi_{L}$ and $\Phi_{LR}$  we, accordingly to Eq.~(\ref{15}) reconstruct a two-particle wave function $\Phi^{(2)}_{R}$ or  $\Phi^{(2)}_{L}$, respectively. 
Therefore, the use of different measurement set-ups allows to explore different bases for a two-particle wave function representation.

\subsection{Distribution functions}
\label{eser}

Electrons emitted by the source on the top of the Fermi sea at low temperatures, Eq.~(\ref{ltl}), have energy larger than the Fermi energy $\mu$. 
It is convenient to count energy $E$ from $\mu$ and to introduce the Floquet energy $ - \hbar\Omega <  \epsilon < 0$ and $\epsilon_{n} = \epsilon + n\hbar\Omega$ with integer $n$.  
Then any energy $E > \mu$ can be represented as $E = \mu + \epsilon_{n}$ with some $n \geq 1$. 
For any function of one energy, $X(E)$, and two energies, $Y(E,E^{\prime})$, let us use the following notations, $X(\epsilon_{n}) \equiv X(E)$  and $Y(\epsilon_{r}, \epsilon^{\prime}_{s}) \equiv Y(E,E^{\prime})$, where  $n$, $r$, $s$ are some integers.

\subsubsection{Single-particle distribution function}

For adiabatic emission we use Eq.~(\ref{fourier}) in Eq.~(\ref{singledistr}) and at low temperatures arrive at the following equation for a single-particle distribution function for electrons ($\epsilon_{n} > 0$),\cite{Moskalets:2006gd}

\begin{eqnarray}
f_{ad}(\epsilon_{n}) = \sum_{m = 0}^{\infty} \left | S_{n+m} \right |^{2}
\,.
\label{sped}
\end{eqnarray}
\ \\
\noindent  
Remind the Fourier coefficients of the scattering matrix, $S_{n+m}$, are calculated at the Fermi energy $\mu$. 

\paragraph{Single-electron emission.} 

If the single particle source $\alpha$ works alone, we have to use $S=S_{\alpha}$. 
Calculating the Fourier coefficients for the function $S_{\alpha}(\tau)$ given in Eq.~(\ref{08}) and substituting them into Eq.~(\ref{sped}) one can find,\cite{Keeling:2006hq,Moskalets:2013kj} 

\begin{eqnarray}
f_{ad,\alpha}(\epsilon_{n}) = 2\Omega\Gamma_{\alpha} e^{ - 2n\Omega\Gamma_{\alpha} }
\,.
\label{sped1}
\end{eqnarray}
\ \\
\noindent
The mean energy of an emitted electron (counted from the Fermi energy) is, 

\begin{eqnarray}
\left\langle \epsilon \right\rangle_{\alpha} \equiv \sum_{n=1}^{\infty} \epsilon_{n} f_{ad,\alpha}(\epsilon_{n}) = \frac{\hbar  }{ 2\Gamma_{\alpha} }
\,.
\label{mean1}
\end{eqnarray}
\ \\
\noindent
This is compatible with dc heat calculations.\cite{Moskalets:2009dk} 
Note in above equation we neglected $|\epsilon| \sim \hbar\Omega$ compared to the rest, since $1/\Gamma_{\alpha} \gg \Omega$. 

Alternatively, given above distribution function $f_{ad,\alpha}(\epsilon_{n})$ can also be calculated via the Fourier transform of the envelope function.
Using Eq.~(\ref{09}) we find 

\begin{eqnarray}
f_{ad,\alpha}(\epsilon_{n}) &=& v_{\mu} {\cal T} \left | A_{\alpha,n} \right |^{2} 
\,,
\nonumber \\
A_{\alpha,n} &=& \int_{0}^{\cal T} \frac{d\tau }{{\cal T} } A_{\alpha}(\tau) e^{i n \Omega \tau }
\,.
\label{sped1wf}
\end{eqnarray}
\ \\
\noindent
Above relation tells us that the mean energy $\left\langle \epsilon \right\rangle_{\alpha}$ can be directly expressed in terms of the wave function $\Phi_{\alpha}$ (or in terms of the envelop function $A_{\alpha}$) as follows,

\begin{eqnarray}
\left\langle \epsilon \right\rangle_{\alpha} &=& 
\int dx \Phi_{\alpha}^{*}(xt) \left [ i\hbar \frac{\partial  }{ \partial t } - \mu \right] \Phi_{\alpha}(xt)
\nonumber \\
&=& 
\int dx A_{\alpha}^{*}(\tau) \left [ i\hbar \frac{\partial  }{ \partial \tau }  \right] A_{\alpha}(\tau)
\,,
\label{mean1wf}
\end{eqnarray}
\ \\
\noindent
where the expression in the square brackets is nothing but the energy operator for emitted particles.

\paragraph{Two-electron emission.}

For a two-particle source composed of two capacitors, $S = S_{L}S_{R}$, with $S_{\alpha}$ given in Eq.~(\ref{08}) we calculate from Eq.~(\ref{sped}),

\begin{eqnarray}
f_{ad}(\epsilon_{n}) = \frac{2\Omega \left( \Delta\tau^{2} + \Gamma_{\Sigma}^{2} \right) }{\Delta\tau^{2} + \Delta\Gamma^{2}  } 
\bigg\{ 
\Gamma_{L} e^{-2n\Omega\Gamma_{L}}
+
\Gamma_{R} e^{-2n\Omega\Gamma_{R}}
 \nonumber \\
\label{16} \\
- \, \frac{4\Gamma_{L} \Gamma_{R} e^{- n \Omega \Gamma_{\Sigma}} }{\Delta\tau^{2} + \Gamma_{\Sigma}^{2} }
\left [ \Gamma_{\Sigma} \cos\left( n\Omega\Delta\tau \right) + \Delta\tau \sin\left(n\Omega\Delta\tau  \right) \right]
\bigg\} \,.
\nonumber
\end{eqnarray}
\ \\
\noindent 
Here $\Delta\tau = \tau^{-}_{R} - \tau^{-}_{L}$, $\Delta \Gamma = \Gamma_{R} - \Gamma_{L}$, and $\Gamma_{\Sigma} = \Gamma_{R} + \Gamma_{L}$. 
Remind that $\tau^{-}_\alpha = t^{-}_{\alpha} - x_{\alpha}/v_{\mu}$ is a reduced emission time, which accounts for a position $x_{\alpha}$ of the source $\alpha$. 
Above function is normalized as follows,

\begin{eqnarray}
\sum\limits_{n=1}^{\infty} f_{ad}(\epsilon_{n}) = 2\,,
\label{normspdf}
\end{eqnarray}
\ \\
\noindent
indicating that there are altogether two electrons (emitted per period) in the state of interest.

If the two sources would work independently they would emit a particle stream which is characterized by the distribution function $\tilde f_{ad}(\epsilon_{n}) = f_{ad,L}(\epsilon_{n}) + f_{ad,R}(\epsilon_{n})$. 
This is the asymptotics of Eq.~(\ref{16}) when two emitted particles do not fill each other, i.e., do not overlap, $|\Delta\tau| \gg \Gamma_{\Sigma}$. 
In contrast, at closer emission times, $|\tau_{R}^{-} - \tau_{L}^{-} | \sim \Gamma_{\Sigma}$, the  wave functions orthogonalization results in an increase of the energy of emitted particles. 
For instance the mean energy $\left\langle \epsilon \right\rangle$ of two emitted particles  exceeds the sum $\left\langle \epsilon \right\rangle_{\Sigma} = \left\langle \epsilon \right\rangle_{L} + \left\langle \epsilon \right\rangle_{R}$. 
Calculating $\left\langle \epsilon \right\rangle$ with the help of the distribution function $f_{ad}(\epsilon_{n})$, Eq.~(\ref{16}), we find

\begin{eqnarray}
\frac{\left\langle \epsilon \right\rangle }{\left\langle \epsilon \right\rangle_{\Sigma} } = 1 +\left | J(\Delta\tau) \right |^{2} \,,
\label{17}
\end{eqnarray}
\ \\
\noindent 
where $J =  \int dx \Phi_{L}^{*}(xt) \Phi_{R}(xt) = 2i\sqrt{\Gamma_{L}\Gamma_{R}}/(\Delta\tau + i \Gamma_{\Sigma})$ is the overlap integral of the wave-functions $\Phi_{L}$ and $\Phi_{R}$, see Eq.~(\ref{09}). 
The same overlap integral appears\cite{Dubois:2013fs} in the problem of the shot noise suppression for electrons colliding at the quantum point contact\cite{Olkhovskaya:2008en,Moskalets:2011jx,Moskalets:2013dl}. 
Note that the mean energy increase shown in Eq.~(\ref{17}) agrees with an enhanced DC heat production of the two-particle emitter.\cite{Moskalets:2009dk}

Since the two sources work jointly, two emitted particles are in states with wave functions $\Phi_{\alpha}$ and $\Phi_{\alpha \bar \alpha}$, see Eq.~(\ref{double}), rather than with $\Phi_{L}$ and $\Phi_{R}$. 
Correspondingly, the distribution function given in Eq.~(\ref{16}) can be represented as the sum of two contributions, 

\begin{eqnarray}
f_{ad}(\epsilon_{n}) = f_{ad,\alpha}(\epsilon_{n}) + f_{ad,\alpha \bar \alpha}(\epsilon_{n})
\,.
\label{decomp}
\end{eqnarray}
\ \\
\noindent 
The first one, $f_{ad,\alpha}(\epsilon_{n})$ is due to a particle in the state with the wave function $\Phi_{\alpha}(xt)$. 
It is given in Eqs.~(\ref{sped1}) and (\ref{sped1wf}). 
The second one, $f_{ad,\alpha  \bar \alpha}(\epsilon_{n})$, is due to a particle in the state with the wave function $\Phi_{\alpha\bar\alpha}(xt)$, Eq.~(\ref{11}).  
It can be calculated by analogy with Eq.~(\ref{sped1wf}) as follows,

\begin{eqnarray}
 f_{ad,\alpha \bar \alpha}(\epsilon_{n}) = 
 v_{\mu} {\cal T} 
 \left | 
 \int\limits_{0}^{{\cal T}} \frac{d\tau }{{\cal T} } 
 A_{\alpha \bar \alpha}(\tau) 
 e^{in\Omega \tau} 
 \right |^{2}  
\,.
\label{sped1wf1}
\end{eqnarray}
\ \\
\noindent
For instance, for $\Delta \tau = 0$ and $\Delta\Gamma=0$ we have,

\begin{eqnarray}
f_{ad,\alpha \bar \alpha}(\epsilon_{n}) 
= 2\Omega\Gamma \left( 1 - 2 n\Omega\Gamma \right)^{2} \, e^{-2n\Omega\Gamma}
\,.
\label{sped2}
\end{eqnarray}
\ \\
\noindent
where $\Gamma = \Gamma_{L} = \Gamma_{R}$.
The energy-dependent prefactor in above equation is formally responsible for increase of the mean energy per particle, see Eq.~(\ref{17}). 
Using $f_{ad,\alpha \bar \alpha}(\epsilon_{n})$ we find the mean energy of an electron to be $\left\langle \epsilon \right\rangle_{\alpha\bar\alpha} = 3\hbar/(2\Gamma)$, compare to Eq.~(\ref{mean1}). 
The energy-dependent prefactor also modifies energy fluctuations $\left\langle \delta^{2}\epsilon \right\rangle = \left\langle \epsilon^{2} \right\rangle - \left\langle \epsilon \right\rangle^{2}$ of emitted electrons.\cite{Battista:2013ew} 
Using Eq.~(\ref{sped1}) and (\ref{sped2}) we find correspondingly,

\begin{eqnarray}
\left\langle \delta^{2}\epsilon \right\rangle_{\alpha} = \left( \frac{\hbar  }{ 2\Gamma} \right)^{2}
\,,
\quad
\quad
\left\langle \delta^{2}\epsilon \right\rangle_{\alpha\bar\alpha} = 5\left( \frac{\hbar  }{ 2\Gamma } \right)^{2}
\,.
\label{enfl}
\end{eqnarray}
\ \\
\nonumber
The absolute value of fluctuations increases for an electron in the state $\Phi_{\alpha\bar\alpha}$ compared to that of an electron in the state $\Phi_{\alpha}$. 
However the relative strength of fluctuations, i.e., compared to the mean energy, decreases~:
$\left\langle \delta^{2}\epsilon \right\rangle_{\alpha}/\left\langle \epsilon \right\rangle_{\alpha}^{2} = 1$ while $\left\langle \delta^{2}\epsilon \right\rangle_{\alpha\bar\alpha}/\left\langle \epsilon \right\rangle_{\alpha\bar\alpha}^{2} = 5/9 < 1$.


The decomposition given in Eq.~(\ref{decomp}) depends on the basis used, $\alpha=L$ or $\alpha = R$. 
Therefore, one cannot attribute any definite distribution function to a single electron, only to two of them together. 
Generally this is due to indistinguishability of particles caused by the overlap of their original wave functions, $\Phi_{R}$ and $\Phi_{L}$, and a subsequent orthogonalization of their actual wave functions, $\Phi_{R}$ and $\Phi_{RL}$ (or $\Phi_{L}$ and $\Phi_{LR}$).   
However, there is also a particular reason why the energy distribution for a single particle is not well defined. 
It is so since the unitary rotation from one basis to another one is time-dependent. 
Hence the energy distributions for basis wave functions are changed during rotation making meaningless the question about energy properties of a separate electron.  
In the limit when two electrons are emitted with a long time delay, $|\tau_{R}^{-} - \tau_{L}^{-}| \gg \Gamma_{\Sigma}$, the two bases converge to each other and the emitted particles become distinguishable. 
In this case one can say which electron is characterized by which distribution function, $f_{L}(E)$ or $f_{R}(E)$. 

All four distribution function, $f_{L}$, $f_{R}$, $f_{LR}$, and $f_{RL}$, can be accessed experimentally using the energy resolved DC current measurement, see Sec.~\ref{dcc}, with one or two sources being switched on by analogy with what is sketched in Sec.~\ref{tbhtm} for a  wave function measurement.

\subsubsection{Two-particle distribution function}

Let us use the adiabatic approximation, Eq.~(\ref{fourier}), and represent the two-particle distribution function, Eq.~(\ref{slater}), with $E = \mu + \epsilon_{r}$ and $s = r + \ell$ as follows,\cite{Moskalets:2006gd}

\begin{eqnarray}
f_{ad}(\epsilon_{r},\epsilon_{s}) = \frac{1 }{2 } \sum_{p=0}^{\infty}\sum_{q=0}^{\infty}
\left | 
\det 
\left(
\begin{array}{ll}
S_{r+p}
&
S_{r+q}
\\
S_{s+p}
&
S_{s+q}
\end{array}
\right)
 \right |^{2} 
 \,.
 \label{18}
\end{eqnarray}
\ \\
\noindent
Remind here all the scattering matrix elements are calculated at the Fermi energy, $E = \mu$.  
In above equation the low temperature limit, Eq.~(\ref{ltl}), is taken. 
It is supposed $\epsilon_{r} > 0$ and $\epsilon_{s} > 0$ since the electronic excitations above the Fermi sea are of interest here.  

\paragraph{Single-electron emission.} 

If only one source $\alpha$ works, we use $S=S_{\alpha}$. 
Taking into account  Eq.~(\ref{08}) we find the corresponding Fourier coefficients, $S_{\alpha,n>0} = - 2\Omega\Gamma_{\alpha} e^{- n \Omega\Gamma_{\alpha}} e^{i n \Omega t_{\alpha}^{-}}$. 
Direct substitution into Eq.~(\ref{18}) gives, $f_{ad}(\epsilon_{r},\epsilon_{s}) = 0$ ($r>0$, $s>0$), as it should be for a genuine single-particle state. 
From Eq.~(\ref{delf1}) we find in this case, $\delta f(\epsilon_{r},\epsilon_{s}) = - f(\epsilon_{r})f(\epsilon_{s})$.

\paragraph{Two-electron emission.} 

In the case when two sources work $S(\tau) = S_{L}(\tau)S_{R}(\tau)$. 
With $S_{\alpha}(\tau)$ from Eq.~(\ref{08}) one can calculate,

\begin{eqnarray}
f_{ad}^{(2)}(\epsilon_{r},\epsilon_{s}) = \frac{8\Omega^{2} \Gamma_{L} \Gamma_{R} \left( \Delta\tau^{2} + \Gamma_{\Sigma}^{2} \right) }{\Delta\tau^{2} + \Delta\Gamma^{2}  } e^{- \Omega\Gamma_{\Sigma}(r+s)}
 \nonumber \\
\times  
\left\{ 
\cosh\left( [r-s]\Omega\Delta\Gamma \right) 
-
\cos\left( [r-s]\Omega\Delta\tau \right)
 \right\} 
 \,.
\label{19} 
\end{eqnarray}
\ \\
\noindent 
Remind $\epsilon_{r} = \epsilon + r \hbar\Omega$ with integer $r \geq 1$ and  the Floquet energy $-\hbar\Omega_{0} < \epsilon < 0$; $\Delta \Gamma = \Gamma_{R} - \Gamma_{L}$ is the difference of coherence times of two sources;  $\Delta \tau = \tau_{R}^{-} - \tau_{L}^{-}$ is the difference of (reduced) emission times, $\tau_{\alpha}^{-} = t_{\alpha}^{-} - x_{\alpha}/v_{\mu}$, with $t_{\alpha}^{-}$  an emission time and $x_{\alpha}$  a position of the source $\alpha$. 

Alternatively above function can be calculated via the double-Fourier transform of the (envelope of the) two-particle wave function, Eq.~(\ref{15}):

\begin{eqnarray}
f_{ad}^{(2)}(\epsilon_{r},\epsilon_{s}) = 
v_{\mu}^{2}{\cal T}^{2} 
\left |
\iint\limits_{0}^{{\cal T}} 
\frac{d\tau d\tau^{\prime} }{{\cal T}^{2} } 
 e^{ir\Omega \tau} 
 e^{is\Omega \tau^{\prime}} 
A_{\alpha}^{(2)} \left( \tau, \tau^{\prime} \right) 
\right |^{2} .
\nonumber \\
\label{tpdf}
\end{eqnarray}
\ \\
\noindent
for either $\alpha=L$ or $\alpha=R$. 

If we sum up  $f_{ad}^{(2)}(\epsilon_{r},\epsilon_{s})$, Eq.~(\ref{19}), over one energy, $\epsilon_{s}$ or $\epsilon_{r}$, then we arrive at the single-particle distribution function, Eq.~(\ref{16}), either $f_{ad}(\epsilon_{r})$ or $f_{ad}(\epsilon_{s})$, respectively, e.g., $\sum_{r=1}^{\infty} f_{ad}^{(2)}(\epsilon_{r},\epsilon_{s}) = f_{ad}(\epsilon_{s}) $.
If we put $r=s$ in Eq.~(\ref{19}) we find $f_{ad}^{(2)}(\epsilon_{r},\epsilon_{r}) = 0 $ as it should be, since two electrons cannot be found in the same state (i.e., with the same energy).   

How to measure a two-particle distribution via an energy-resolved shot-noise was discussed in Sec.~\ref{zfn}. 
Here let us estimate a feasibility of such a proposal for an adiabatic emission regime. 
The energy scale, over which the distribution function, Eq.~(\ref{19}),  changes, is $\sim \hbar/(\Gamma_{L} + \Gamma_{R})$. 
For the adiabatic regime of the source of Ref.~\onlinecite{Feve:2007jx} the coherence time $\Gamma_{\alpha} \sim T_{\alpha}/(2\pi \Omega)$, where $T_{\alpha}$ is the transmission probability of the quantum point contact connecting the capacitor and the waveguide.\cite{Olkhovskaya:2008en}  
At $T_{\alpha} \sim 0.2$ and $\Omega \sim 2\pi \cdot 500~{\rm MHz}$ we find $\Gamma_{\alpha} \sim 10~{\rm ps}$. 
The width $\gamma$ of the resonance level of an energy filter should satisfy the following inequality, $\hbar\Omega \ll \gamma \ll \hbar/(2\Gamma_{\alpha})$. 
For the parameter chosen it  becomes (in temperature units), $24~{\rm mK} \ll \gamma/k_{B} \ll 380~{\rm mK}$. 
The energy filter used in Refs.~\onlinecite{Altimiras:2010ej,leSueur:2010fg,Altimiras:2010dk} is characterized by $\gamma/k_{B} \sim 50~{\rm mK}$. 
That is quite reasonably for the purposes we are discussing.

\section{Conclusion}
\label{c}

I analyzed a two-particle state emitted by two uncorrelated but synchronized single-electron sources (e.g., periodically driven quantum capacitors) coupled in series to the same chiral electronic waveguide. 
The two-particle correlation function for emitted state is expressed in terms of the Floquet scattering matrix of a combined two-particle source. 
The Fourier transform of the correlation function, the two-particle distribution function, is calculated and related to a cross-correlation function of currents flowing  through the energy filters, quantum dots with a single conductive level each, Fig.~\ref{set-up}. 
   
In the case of emitters working in the adiabatic regime, the two-particle wave function is calculated and represented in two equivalent but different forms depending on the single-particle wave functions used as a basis, see Fig.~\ref{source}. 
The existence of these two bases is rooted in the presence of two single-particle emitters, which affect each other. 
Let denote as $\Psi_{L}$ and $\Psi_{R}$ the wave functions of a single electron emitted by one or another source if they would work independently. 
The presence in a waveguide of an electron emitted by one source affects the emission of an electron by the other source such that the actual single-particle wave functions become orthogonal and hence cannot be just $\Psi_{L}$ and $\Psi_{R}$, which in general are not orthogonal. 
The simplest way to construct orthogonal single-electron wave functions is to take one of them, say $\Psi_{L}$, unperturbed and to orthogonalize the other, denote it as $\Psi_{LR}$. 
Alternatively one can keep unperturbed $\Psi_{R}$ and orthogonalize the other, $\Psi_{RL}$. 
These two bases, $\Psi_{L}$, $\Psi_{LR}$ and $\Psi_{R}$, $\Psi_{RL}$  are exactly what appears naturally when the first-order correlation function for the state emitted by the two-particle source  is considered, see Eqs.~(\ref{10}) and (\ref{double}). 
In particular, when the electrons are emitted with a long time delay such that they do not overlap, the wave functions $\Psi_{LR}$ and $\Psi_{RL}$ approach to $\Psi_{R}$ and $\Psi_{L}$, respectively.      
What important is that in the general case all four single-electron wave functions are accessible  experimentally.

\begin{acknowledgments}
I am grateful to Markus B\"{u}ttiker for initiating this work and numerous helpful discussions. 
I thank Christian Glattli and Francesca Battista for useful discussions and comments on the manuscript. 
I appreciate the warm hospitality of the University of Geneva where this work was started. 
\end{acknowledgments}

\end{document}